\documentclass[twocolumn,preprintnumbers,amsmath,amssymb]{revtex4}

\usepackage{graphicx}
\usepackage{dcolumn}
\usepackage{bm}
\usepackage{subfigure}

\newcommand{\bra}[1]{\langle{#1}|} 
\newcommand{\ket}[1]{|{#1}\rangle} 

\begin{document}

\title{Lifetime measurements of the \(5d\) states of rubidium}

\author{D. Sheng}
\author{A. {P{\'e}rez Galv{\'a}n}}
\author{L. A. Orozco}
\affiliation{Joint Quantum Institute, Department of Physics,
University of Maryland, and National Institute of Standards and
Technology, College Park, MD 20742, U. S. A.}

\begin{abstract}

We present lifetime measurements of the $5D_{3/2}$ and $5D_{5/2}$
states of rubidium using the time correlated single photon
counting method. We perform the experiment in a magneto-optical
trap of $^{87}$Rb atoms using a two-step excitation with the trap
laser at 780~nm as the first step. We record the 761.9~nm
fluorescence from the decay of the $5D_{3/2}$ state to the
$5P_{1/2}$ state, and measure the lifetime of the $5D_{3/2}$ state
$\tau=246.3(1.6)$~ns. We record the 420.2~nm fluorescence from the
cascade decay of the $5D_{5/2}$ state to the $5S_{1/2}$ state
through the $6P_{3/2}$ state, and extract the lifetime of the
$5D_{5/2}$ state ${\tau}=238.5(2.3)$~ns.

\end{abstract}

\maketitle

\section{Introduction}

Lifetime measurements are fundamental for the understanding of
atomic structure. The precisely measured quantities of transition
energies and excited state lifetimes are important tests of atomic
structure theory through comparisons with the values derived from
the calculated energies and transition matrix elements
\cite{safronova08}. Alkali atoms and alkali-like ions are the
benchmark systems where measurements and calculations have reached
the highest development because the simple single-core electron
structure facilitates the application of advanced calculation
methods and current simple laser cooling techniques.

A large number of lifetime measurements exist in the alkali atoms
and alkali-like ions for the $s$ and $p$ levels, but there are not
as many high precision measurements on the $d$ and higher angular
momentum states. These states are becoming important not only in
the study of fundamental symmetries
\cite{ginges04,fortson93,dzuba01,kreuter05}, but also in quantum
information science as they are used for qubit manipulation in ion
traps \cite{steane97,nagerl00}.

The $d$ states of rubidium are interesting from the calculation
point of view as they remain in the domain of non-relativistic
physics, but they are thoroughly affected by correlation
corrections. Safronova {\it et al.} \cite{safronova04} have shown
the important role of high order corrections, up to third order,
in calculations that use many body perturbation theory (MBPT) of
lifetimes of these states. We present in this paper a precise
measurement of the $5D_{3/2}$ and $5D_{5/2}$ states of rubidium.
Our group has studied the equivalent states in francium, the
heaviest alkali atom, and achieved precisions of  0.4\% and 4.3\%
for the $7D_{3/2}$ and $7D_{5/2}$ states, respectively
\cite{grossman00b}.

Previous experimental work on the lifetime of the Rb $5D_{3/2}$
state \cite{tai75} and the whole $5d$ manifold \cite{marek80}
achieved a precision long surpassed by atomic calculations. This
paper shows an improvement by more than a factor of twenty in the
precision of the $5D_{3/2}$ state lifetime measurement and a
precise determination of the $5D_{5/2}$ state lifetime. These
improvements will hopefully trigger improved calculations as the
precision of this work ($\sim1\%$) is better than the current
estimates in the theory ($\sim5\%$) \cite{safronova04}.

This paper has five sections. Sec.~\ref{sec:theory} presents the
theoretical background. Sec.~\ref{sec:setup} explains the
experimental method and apparatus. Sec.~\ref{sec:5d23} and
\ref{sec:5d25} presents the data analysis for the measurements of
the $5D_{3/2}$ and $5D_{5/2}$ state lifetimes, respectively.
Sec.~\ref{sec:end} gives a conclusion.

\section{Theoretical background\label{sec:theory}}

Suppose an atom is in an excited state $\ket{e}$, and it decays to
a set of lower states $\{\ket{l_j}\}$, the lifetime of this
excited state, $\tau_e$, is \cite{safronova04}
\begin{equation}
\frac{1}{\tau_e}=\sum_{j}\frac{4}{3}\frac{\omega^3_{el_j}}{c^2}\alpha\frac{|\bra{J_e}|\bm{r}|\ket{J_{l_j}}|^2}{2J_e+1},
\label{eq:theory}
\end{equation}
where $c$ is the speed of light, $\hbar\omega_{el_j}$ is the
transition energy between $\ket{e}$ and $\ket{l_j}$, $\alpha$ is
the fine structure constant, $J$ is the angular momentum
associated with each state, and $\bra{J_e}|\bm{r}|\ket{J_{l_j}}$
is the reduced transition matrix element.

Equation~(\ref{eq:theory}) shows that lifetime measurements probe
the overlap of wavefunctions of electron states with different
quantum numbers at large distance. This is in contrast with
precise energy measurements which probe the overall distribution
of electron wavefunction, or the hyperfine splittings that are
sensitive to the short range of the electronic wavefunctions.

Theoretical calculations aim to first find out the correct set of
eigenfunctions of the atomic system Hamiltonian, and then
calculate the physical properties using those results. In
relativistic many-body calculations, there are different methods
to obtain the eigenfunctions of the system. The most
straightforward way is to start with a Dirac-Hartree-Fock (DHF)
wavefunction ($\ket{\psi^{(0)}}$) and iterate the perturbation
theory order by order. This is the relativistic many-body
perturbation theory. This method usually stops at the third order
due to the large calculation load. This limits the precision of
the results that can also suffer when the mean field ground state
is not a good starting point. An improvement is to use a
wavefunction including some excitations to the DHF wavefunction
\cite{safronova08}:
\begin{equation}
\ket{\psi}=(1+S_1+S_2+S_3+\ldots)\ket{\psi^{(0)}},\label{eq:mbpt}
\end{equation}
where $\ket{\psi^{(0)}}$ is the DHF wavefunction and $S_i$ is the
$i$-body excitation operator. The method involving only $S_1$ and
$S_2$ operators in Eq.~(\ref{eq:mbpt}) is the singles-doubles (SD)
method. The so-called all-order method includes $S_1$, $S_2$ and
the dominant parts of the $S_3$ (see Ref.~\cite{safronova99} for
an extended explanation of the method).

The calculation of transition matrix elements limits the precise
prediction of the lifetimes. For the all-order method, this
problem becomes serious for some levels with large correlation
effects (such as $d$ levels), where the wavefunction in the form
of Eq.~(\ref{eq:mbpt}) is a bad starting point of calculations.
Theorists use a semi-empirical scaling scheme for certain
wavefunction coefficients to solve the problem. For the rubidium
$5d$ states, the lifetimes calculated with and without scaling
have a relative difference of 20\%, but it is hard to assert the
correctness of the scaling due to high uncertainties of past
experimental results \cite{safronova04}.

Lifetime measurements help on the extraction of transition matrix
elements. If there is only one decay channel, we can extract the
absolute value of the reduced transition matrix element directly
from the lifetime data. For example, by measuring the lifetimes of
the excited states, Ref.~\cite{simsarian98} extracts the
transition matrix elements $|\bra{7P_{1/2}}|r|\ket{7S_{1/2}}|$ and
$|\bra{7P_{3/2}}|r|\ket{7S_{1/2}}|$ of Fr, and
Ref.~\cite{hoeling96} extracts the transition matrix element
$|\bra{5D_{5/2}}|r|\ket{6P_{3/2}}|$ of Cs. The case of a
multi-channel decay sometimes requires a combination of various
experimental approaches to extract the transition matrix elements
\cite{safronova08}, including the measurement of static scalar
polarizability \cite{Arora07}, and nonresonant two-photon,
two-color linear depolarization spectrum \cite{bayram00}.

\section{Experimental method and setup\label{sec:setup}}

We adapt the time correlated single photon counting method
\cite{oconnor84} in this experiment with a cold sample of
$^{87}$Rb atoms (temperature less than 100 $\mu$K) in a
magneto-optical trap (MOT). After exciting the atoms to the $5d$
states, we turn off the excitation beams and record the delay
between the fluorescence and a fixed time reference. Here, the
single photon counting technique refers to the fact that we record
at most one photon in one duty cycle. It is possible for the
detector to detect more than one photon in one cycle, but once the
electronics record one signal, they will not accept another until
a new cycle begins. This method works best when the probability of
detecting more than one photon in one cycle is very small, which
in turn keeps the corrections low (see the pulse pileup
corrections in the systematic study of Sec. \ref{sec:5d23}). In
our experiments, we detect one photon in about one hundred cycles.

We use a rubidium dispenser as the atomic vapor source and the MOT
resides inside a 15~cm radius spherical chamber with the vacumm
pressure of $10^{-10}$ Torr. A pair of anti-Helmholtz coils
provides a magnetic gradient of 6 G/cm and three pairs of
Helmholtz coils provide the fine tuning of the magnetic
environment. A Coherent 899-01 Ti:Sapphire laser with linewidth
better than 100 kHz provides three pairs of MOT trapping beams
with intensity of 8 mW/cm$^2$, and the laser is red detuned from
the $5S_{1/2},F=2$ $\rightarrow$ $5P_{3/2},F=3$ transition by
approximately 20 MHz. A Toptica SC110 laser provides the repumper
beam with intensity of 3 mW/cm$^2$ and it is on resonance with the
transition $5S_{1/2},F=1$ $\rightarrow$ $5P_{3/2},F=2$. We capture
about $10^5$ atoms in the MOT with diameter of 600 $\mu$m and peak
density of around $10^9$ cm$^{-3}$. We use two charge-coupled
device (CCD) cameras to monitor the fluorescence of the MOT in two
perpendicular directions.

\begin{figure}
\includegraphics[width=8cm]{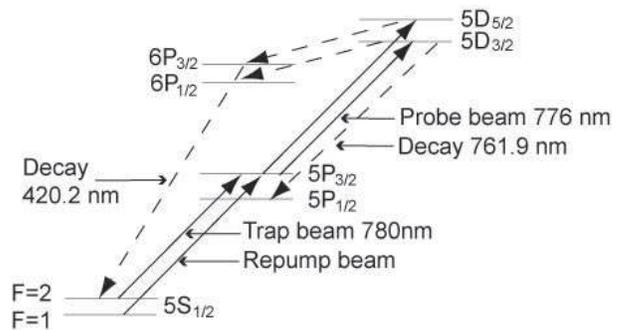}
\caption{\label{fig:level} $^{87}$Rb energy levels involved in the
experiment.}
\end{figure}

We list the relevant energy levels of $^{87}$Rb for this
experiment in Fig.~\ref{fig:level}. We use a two-step transition
to reach the $5d$ states, where the trapping beam of the MOT
enables the first step and the $5P_{3/2},F=3$ state is the
intermediate state. A SDL TC40-D laser with linewidth of 5 MHz
provides the probe beam to reach the 5$d$ states. We inject the
probe beam to the MOT region through a single mode fiber, which
sets the waist ($1/e^2$ power) to 1.2 mm. The power of the probe
beam is 0.5 mW for the excitation to the $5D_{5/2},F=4$ state, and
1.0 mW for the excitation to the $5D_{3/2},F=3$ state.

We lock the frequency of the trapping beam using the
Pound-Drever-Hall method with saturation spectroscopy in a Rb
cell. We send part of the frequency modulated light employed on
this lock to an independent rubidium glass cell, where this light
overlaps with a small laser beam taken from the probe beam. We
monitor the absorption of the 780~nm light as a function of the
frequency of the 776~nm laser, and the intermodulation of the
sidebands yields error-signal like features that we use to lock
the frequency of the probe beam on resonance
\cite{perezprep,boucher92}.

\begin{figure}[hpt]
\includegraphics[width=6cm]{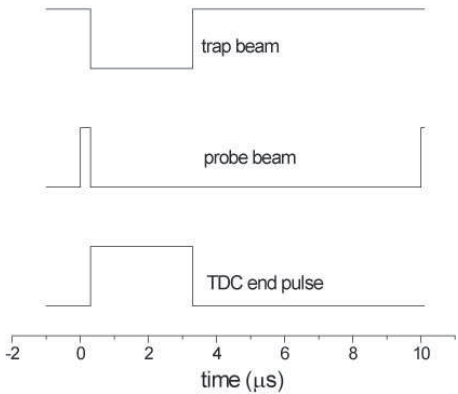}
\caption{\label{fig:time} Time sequences in the lifetime
measurement. Each cycle has a period of 10 $\mu$s.}
\end{figure}

We use a cycle of 10 $\mu$s length, and have two different schemes
for photon detection and time control for the two different
measurements. In the $5D_{3/2}$ state lifetime measurement, we put
a 760~nm interference filter with bandwidth of 10~nm (Andover
760FS10-25) in front of the detector, a Hamamatsu R636
photomultiplier tube (PMT) with quantum efficiency of 10\% at this
wavelength. Since a large amount of 780~nm photons from the
scattered trapping beam pass through the filter, we turn the
trapping beam off after the excitation phase to decrease the
background. We use two acousto-optical modulators (AOM) to turn on
and off the trapping beam and the probe beam. In the $5D_{5/2}$
state lifetime measurement, we use a 420~nm interference filter
with a bandwidth of 10~nm (Andover 420FS10-25) in front of the
detector, a PerkinElmer C962 channel PMT with quantum efficiency
of 15\% at this wavelength. The trapping beam has a negligible
contribution to the background in this case. We use a Gs\"{a}nger
LM0202P electro-optical modulator (EOM) and an AOM to chop the
probe beam for this experiment. The turn off ratio of EOM and AOM
is better than 30:1 in 30~ns. Fig.~\ref{fig:time} shows the time
sequence of the pulses.

\begin{figure}[hpt]
\includegraphics[width=8cm]{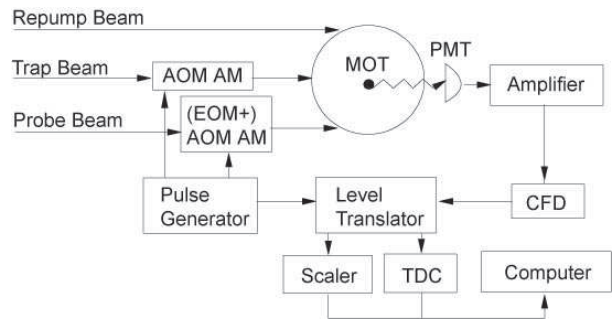}
\caption{\label{fig:detect} The setup of the detection and signal
process.}
\end{figure}

The PMT is about 35 cm away from the MOT and we use a 10X Computar
Macro Zoom lens in front of it for light collection. Two
synchronized Stanford Research DG535 pulse generators, which have
a 5 ps delay resolution and 50 ps rms jitter, provide all the time
references in the signal process. The signal from the PMT goes
through the amplifiers first. In the $5D_{3/2}$ ($5D_{5/2}$)
lifetime measurement, we amplify the signal from the Hamamastsu
R636 PMT (channel PMT) 64 (4) times using an EG{\&}G AN106/n plus
an AN101/n DC amplifier (an AN106/n amplifier alone). The output
goes through an Ortec 583 constant fraction discriminator (CFD)
and a Lecroy 7126 level translator. The level translator converts
the input signal to ECL, TTL and NIM outputs. We send the output
of the NIM signal to a Stanford Research SR430 multi-channel
scaler to monitor the photon counting histogram during the
experiment. We send the ECL signal as a start pulse to a Lecroy
3377 time-to-digital converter (TDC), which has a resolution of
0.5~ns and is triggered by the falling edge of the input pulse.
The TDC measures the delay between the observed photon and the
fixed pulse given by the pulse generator (see Fig.~\ref{fig:time}
and Fig.~\ref{fig:detect}). The output of the TDC goes to a Lecroy
4302 memory and we read out the results through a Lecroy 8901A
GPIB interface. Fig.~\ref{fig:detect} shows the diagram of the
signal process.

\section{Lifetime of the $5D_{3/2}$ state\label{sec:5d23}}
\begin{figure}
\includegraphics[width=7cm]{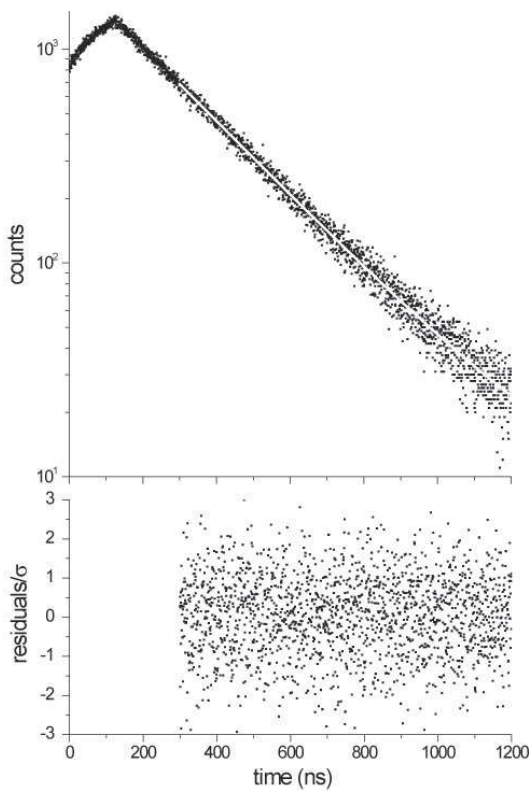}
\caption{\label{fig:counts5D23}Upper plot: time distribution of
761.9~nm photons (black points) in the measurement of the
$5D_{3/2}$ state with the fitting function (white line). Lower
plot: normalized fit residuals. We have subtracted the background
in the decay curve and we use the statistical error of the data to
normalize the fit residuals.}
\end{figure}

We excite the atoms to the $5D_{3/2}$ state and about 50\% of the
atoms decay into the $5P_{1/2}$ state with fluorescence at
761.9~nm, where we get the branching ratios using the transition
matrix elements from Ref.~\cite{safronova04}. We record the
fluorescence and accumulate data until the peak count is more than
1000 for each data set. This usually takes half an hour to 45
minutes. The ratio of peak signal to background is generally
around 40, and we take an additional data set of background for
roughly the same time. The upper plot of Fig.~\ref{fig:counts5D23}
shows a typical data set after substraction of background. There
is another possible way to observe the decay through the detection
of 420~nm fluorescence that comes from the $6p$ manifold (see
Fig.~\ref{fig:level}). This has a great background advantage, but
requires fitting to a sum of three exponentials, which challenges
our fitting process (see explanations in the Sec.~\ref{sec:5d25}).
The results that we present here are consistent with a preliminary
analysis of the blue decay.

\subsection{Decay model}\label{sec:decay}

The data in Fig.~\ref{fig:counts5D23} shows an increase of the
fluorescence due to the excitation of the system by the probe beam
and the decay after turning off the probe beam. We qualitatively
understand the rising part by treating this many-level system as a
four-level system. We denote the ground state, the $5P_{3/2}$
state and the $5D_{3/2}$ state as levels 1, 2 and 3, and treat all
other states that the $5D_{3/2}$ state decays to as level 4. The
excitation rate from level $i$ to $j$ is $A_{ij}$, the spontaneous
decay rate of level $k$ is $R_k$ and the branching ratio of decay
channel from $k$ to $l$ is $\gamma_{kl}$. Because the system is
already in the steady state before we turn on the weak probe beam,
we take the second step transition as a small perturbation to the
system and get the rate equations:
\begin{eqnarray}
&&\dot{N}_3=A_{23}N_2-R_3N_3\nonumber\\
&&\dot{N}_4=\gamma_{34}R_3N_3-R_4N_4. \label{eq:decaymodel}
\end{eqnarray}

The initial condition is $N_1(0)=N_2(0)=N_0/2$, where $N_0$ is the
total number of atoms and the time origin is the time we turn on
the probe beam.  This gives the number of atoms in level 3 as
\begin{equation}
N_3(t)=\frac{A_{23}}{R_3}\frac{N_0}{2}[1-\exp{(-R_3t)}], t<t_1,
\label{eq:5d23rise}
\end{equation}
where $t_1$ is the time we turn off the probe beam. This
qualitatively agrees with the rising behavior of the signal.

The number of atoms in this level begins to decay when we turn off
the probe beam. For a perfect turnoff at time $t_1$, we get the
decaying behavior:
\begin{equation}
N_3(t)=N_3(t_1)\exp{[-R_3(t-t_1)]}, t>t_1. \label{eq:5d23decay}
\end{equation}

The intensity of the fluorescence from the $5D_{3/2}$ state to the
$5P_{1/2}$ state is proportional to the atom number in the
$5D_{3/2}$ state and Eq.~(\ref{eq:5d23decay}) shows that the
decaying part of the signal is independent of the excitation part
except for a coefficient.

\subsection{Fitting process}
We use the Levenberg-Marquardt (L-M) algorithm \cite{press87} and
fit the data to the function
\begin{equation}
y=A{\exp}(-t/\tau)+B, \label{eq:fit5d23}
\end{equation}
where the $\tau$ is the $5D_{3/2}$ state lifetime and B describes
the background. The fitting is a procedure to minimize $\chi^2$:
\begin{equation}
\chi^2=\displaystyle\sum_i\omega_i(y_i-y_f)^2\label{eq:chi2},
\end{equation}
where $y_i$ and $y_f$ are the experimental and fitting value for
the $i$th data point, and $w_i$ is the inverse of the square of
the statistical error of the data assuming the data follow the
Poisson distribution.

We use two criteria on the choices of the range of the data to
fit. First, we choose the starting point of the fitting range away
from the turning off point to avoid pulse turnoff effects. Second,
we need to make sure that the fitting results are stable when we
vary the starting and end points of the fitting range. This
fitting process gives us an average of reduced $\chi^2=1.03$ for
all the data sets we have taken. The lower plot of
Fig.~\ref{fig:counts5D23} shows the normalized fit residuals of
the upper plot data.

\subsection{Statistical error}
We have taken 28 sets of data with the same experimental
conditions, with each individual data set fitting error,
$\sigma_i$, of around 1\%. Considering this sample of 28 data
points, we obtain a weighted mean lifetime of 246.5~ns, with the
error given by $({\sum{1/\sigma_i^2}})^{-1/2}$, which is 0.3~ns.
We choose to scale the uncertainty by a factor \cite{pdg2006}
\begin{equation}
S=\left[{\chi^2/(N-1)}\right]^\frac{1}{2},
\end{equation}
where $\chi^2$ is defined in Eq.~(\ref{eq:chi2}), and $N$ is the
number of data points. In this way, the mean value does not
change, but it helps to solve the problem of underestimation of
the error on some data sets. We get $S=2.1$, and the statistical
error as 0.6~ns.

\subsection{Systematic effects}

We focus on the following systematic effects:

\emph{Time calibration and height nonuniformity of TDC.} We send
two pulses with a fixed delay from the DG535 pulse generator as
the start and end pulses to the TDC. By comparing the readout of
the TDC with the calibrated delay, we get a nonlinearity of the
TDC time calibration of less than 0.01\%, which in turn gives an
error on the lifetime measurement of less than 0.01\%. To check
the height uniformity of the TDC channels, we change the source of
the photon to the scattering of laser light on a paper. We control
the power of the laser so that the recording rate is similar to
that in the lifetime experiments, and accumulate data until the
count of each channel reaches 100. We fit the results with a
linear function (fits to higher order polynomials yield the same
conclusion) and get a slope of 0.1 count per 1000 channels, which
gives an error of 0.01\% on the lifetime measurement.

{\emph{Pulse pileup correction.}} There is a difference between
the number of the signals detected and signals recorded in the
single photon counting method, this leads to pulse pileup
correction. Suppose we have $N_E$ cycles of excitation, and get
$N_j$ counts on the $j$th channel of the TDC, then the pulse
pileup corrected result of the $i$th channel is
\begin{equation}
N_i'=\frac{N_i}{1-\frac{1}{N_E}\displaystyle\sum_{j<i}N_j}.
\end{equation}
This gives us a correction of around -0.1\% for our measurements.

{\emph{Quantum beats.}} Quantum beats come from the interference
of the decay paths from several coherently excited states to the
same lower state. By considering this effect, we rewrite the
Eq.~(\ref{eq:fit5d23}) as \cite{demtroder96}
\begin{equation}
y=A\exp(-t/\tau)[1+\displaystyle\sum_{i}C_i\cos(\omega_i{t})]+B,
\label{eq:qbeat}
\end{equation}
where the sum is over all possible interference paths and
$\hbar\omega_i=\Delta{E_i}$ is the energy difference of two
excited states involved in the process. This has been an important
issue when the probe beam has a large bandwidth \cite{young94}.

We first consider the possibility of the quantum beats from the
hyperfine structure. The energy difference between $5D_{3/2},F=2$
and $F=3$ sates is around 40 MHz \cite{tai75}. This separation is
much larger than the probe laser linewidth. The natural linewidth
of the $5D_{3/2}$ state is 0.6 MHz, considering the power
broadening effect, the transition linewidth is at most 6 MHz. A
sharp turnoff process could generate a large bandwidth in
frequency domain. We decompose the probe pulse in frequency range
by the fast Fourier transform (FFT), and find that the high
frequency components (higher than 40 MHz) only take about $1/200$
of the total power. In conclusion, the probability to coherently
excite the hyperfine states is negligible. The FFT of the fitting
residuals does not show any peaks around the frequency of the
hyperfine splitting either.

We position the atoms at the zero of the magnetic field in the
MOT, but the finite extent of the cloud sample regions with a
magnetic field. The Zeeman splitting introduced by this magnetic
field could be a source of quantum beats, because we always leave
the gradient magnetic field for the MOT on during the experiment.
However, the overlapped $\sigma+$ and $\sigma-$ trapping beams
tend to average out this effect.

We estimate an upper bound of the effect due to the Zeeman
splitting following the treatment in Ref.~\cite{grossman00b}. We
have $\hbar\omega=\Delta{m_F}g_F\mu_B{B}$, where $\Delta{m_F}=2$,
$g_F=0.4$ for $5D_{3/2},F=3$ state, and $\mu_B$ is the Bohr
magneton. The size of the MOT is around 600 $\mu$m and the
magnetic gradient is around 6 G/cm, we use the maximum magnetic
field 0.18 G in that region for the calculation. This gives us a
maximum beat frequency of 202 kHz, which sets an upper bound of
this error of 0.15\%.

\begin{figure}[hpt]
\includegraphics[width=7cm]{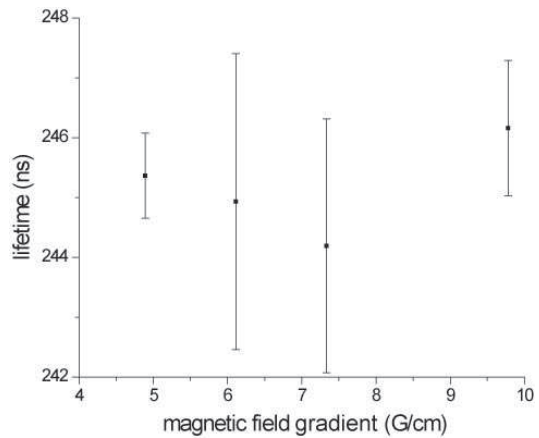}
\caption{\label{fig:mag} The systematic study of the effect of the
magnetic field gradient on the $5D_{3/2}$ lifetime measurement.}
\end{figure}

We also look for possible effects of the magnetic field by
changing the magnetic field gradient from 5 G/cm to 10 G/cm and do
not find any systematic changes. Fig.~\ref{fig:mag} shows the
results, where the smaller error bars in the points with 4.9 G/cm
and 9.8 G/cm are due to more data sets under those conditions in
this systematic study. Studies of linear correlation coefficient
of the data points show that they are highly uncorrelated. We get
an error of 0.16\% by this effect, which is the standard deviation
of the data points taken in this study. We use this measured bound
of 0.16\% for possible magnetic effects that include the quantum
beats.

{\emph{Radiation trapping.}} Radiation trapping comes from the
reabsorption of the emitted photon by the sample itself. This
effect depends on the length and density of the sample and could
increase the measured lifetime substantially when there are many
atoms \cite{gomez05b}. The MOT has the advantage of its size,
which makes this effect much smaller compared with the experiment
in the cell. Denote $l$ and $n$ as the length and density of the
sample, and $\alpha$ as the atomic absorption coefficient.
Ref.~\cite{kibble67} gives an expression for this effect:
\begin{equation}
\frac{\tau'}{\tau}=1+\left(\frac{C}{\lambda}\right)^2,
\label{eq:radtrap1}
\end{equation}
where $C=l {\alpha} n$ and $\lambda\tan(\lambda)=C$. In the limit
of small length, Eq.~(\ref{eq:radtrap1}) becomes
\begin{equation}
(\tau'-\tau)={\tau}C={\tau}l\alpha{n},
\end{equation}
where the change of lifetime depends linearly on density. We have
changed the density of our sample by a factor of six to look for
this effect, and do not observe any systematic changes. The error
due to this effect is smaller than 0.1\%.

\emph{Other systematics.} The pulse length and power of the probe
beam are the parameters controlling the number of atoms in the
excited states. By changing the pulse length from 320~ns to
720~ns, and also the power from 0.5~mW to 2.4~mW, we do not
observe systematic changes. We set the error on those effects as
0.6\%, which we obtain in the same sprit as in the study of the
magnetic field effects.

\subsection{Summary and comparison}\label{subsec:redsum}
We list the error budget of the $5D_{3/2}$ lifetime measurement in
table \ref{tab:err5d23}, and compare our result with the previous
experimental and theoretical work in table \ref{tab:ret5d23}. On
the experimental part, Ref. \cite{tai75} made use of the Hanle
effect to extract the lifetime. In that case, they could not
optically excite the atoms to the $5D_{3/2}$ state, instead, they
optically excited the atoms to a higher $p$ state and populated
the desired level by spontaneous emission. On the theoretical
part, Ref.~\cite{theodosiou84} used the quantum defect theory with
a realistic potential, and Ref.~\cite{safronova04} used the scaled
all-order method.

The result of Ref.~\cite{safronova04} sits on the edge of
2$\sigma$ range of our data and that of Ref.~\cite{theodosiou84}
sits on the edge of the 4$\sigma$ range. For more detailed
comparison, we need to know the range of validity of the
theoretical calculation. A possible way to define the error of the
all-order method calculation is comparing its results with third
order many body perturbation theory. In this way, the error of the
all-order method calculation for the Rb $5d$ states lifetime is
around 5\% \cite{safronovapriv}, and our result of the $5D_{3/2}$
state lifetime stays within its prediction. Our result $1\sigma$
range has a small overlap with the $1\sigma$ range of the result
of Ref.~\cite{tai75}, which has a much bigger error bar.

\begin{table}
\caption{\label{tab:err5d23} Error budget of the $5D_{3/2}$ state
lifetime measurement.}
\begin{ruledtabular}
\begin{tabular}{lcc}
Source&Correction (\%)&Error(\%)\\
\hline
Statistical&&$\pm$0.25\\
Time calibration&&$<\pm$0.01\\
TDC nonuniformity&&$\pm$0.01\\
Pulse pileup&-0.1&\\
Quantum beats and magnetic field&&$\pm$0.16\\
Radiation trapping&&$<\pm$0.10\\
Other Systematics&&$\pm$0.60\\
\\
\textbf{Total}&&$\pm$\textbf{0.66}
\end{tabular}
\end{ruledtabular}
\end{table}

\begin{table}
\caption{\label{tab:ret5d23} Comparison of the measured lifetime
of the $5D_{3/2}$ state with the previous work.}
\begin{ruledtabular}
\begin{tabular}{llc}
&& $\tau_{5D_{3/2}}$ (ns)\\
\hline
Experiment&\textbf{This work}&\textbf{246.3$\pm$1.6}\\
&Tai \emph{et al.} \cite{tai75}&205$\pm$40\\
\\
Theory&Theodosiou \cite{theodosiou84}&240\\
&Safronova \emph{et al.} \cite{safronova04}&243
\end{tabular}
\end{ruledtabular}
\end{table}

\section{Lifetime of the $5D_{5/2}$ state\label{sec:5d25}}

\begin{figure}
\includegraphics[width=7cm]{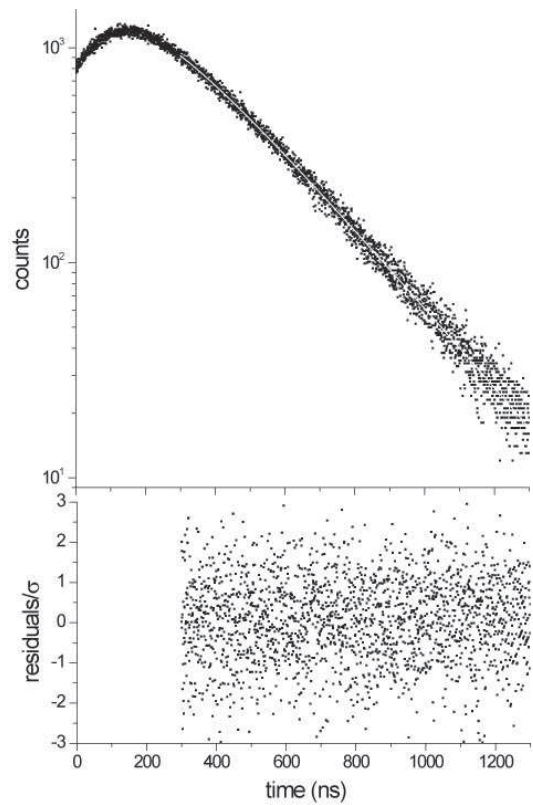}
\caption{\label{fig:counts5D25}Upper plot: time distribution
(black points) of the blue photon in the measurement of the
$5D_{5/2}$ state with the fitting function (white line). Lower
plot: normalized fit residues, where the normalization factor is
the statistical error of the data.}
\end{figure}

About 30\% of the atoms in the $5D_{5/2}$ state decay to the
$6P_{3/2}$ state with fluorescence at 5~$\mu$m, and about 25\% of
the atoms in the $6P_{3/2}$ state decay to the ground state with
fluorescence at 420.2~nm, where we get the branching ratios in the
same way as in Sec.~\ref{sec:5d23}. We record this blue
fluorescence and accumulate data until the peak count is more than
1000 for each data set. The upper plot of
Fig.~\ref{fig:counts5D25} shows a typical data set. Since we
detect the blue photon, the background is very small (less than 1
count per channel in half an hour), and we do not have to take an
additional background data.

Most of the data process and systematic effects studies in this
measurement are similar to those in the measurement of the
$5D_{3/2}$ state, however, there are some new features in the
analysis.

\subsection{Decay model}
We use the same model in the previous section. The differences are
that we change the $5D_{3/2}$ state to the $5D_{5/2}$ state and
specify level 4 as the $6P_{3/2}$ state, because the fluorescence
we observe in this experiment is proportional to number of atoms
in the $6P_{3/2}$ state. Using the same notation as in
Sec.~\ref{sec:decay}, we get the rising and decaying part of the
signal in this experiment as
\begin{eqnarray}
N_4(t)=&&\frac{\gamma_{34}R_3}{R_4-R_3}\frac{A_{23}N_0}{2}\nonumber\\
&&\left[\frac{1-\exp(-R_3t)}{R_3}-\frac{1-\exp(-R_4t)}{R_4}\right],t<t_1\nonumber\\
N_4(t)=&&\left[N_4(t_1)-\frac{\gamma_{34}R_3}{R_4-R_3}N_3(t_1)\right]\exp{[-R_4(t-t_1)]}\nonumber\\
&&+\frac{\gamma_{34}R_3}{R_4-R_3}N_3(t_1)\exp{[-R_3(t-t_1)]},t>t_1\label{eq:5d25deacy}
\end{eqnarray}

This shows that the curve observed in this experiment includes the
information of the lifetimes of the $5D_{5/2}$ state and the
$6P_{3/2}$ state.

\subsection{Fitting process}

The ratio between coefficients of the two exponentials depends on
the lifetimes and initial atom number at the turnoff point of both
states (see Eq.~(\ref{eq:5d25deacy})). If we want to fit the decay
curve according to the function in Eq.~(\ref{eq:5d25deacy}), we
have to evaluate the atom number by doing the convolution of the
excitation pulse with the formula that determines population in
the $6P_{3/2}$ state. A simple way to avoid this process is to
leave both coefficients as free parameters and fit the decay
signal away from the turnoff point \cite{gomez05b}. The existence
of a stable range of fitting results when we change the starting
and end points of the fitting range is a test of the validity of
this treatment.

We use the following fitting function to describe the data
\begin{equation}
y=A_1\exp(-t/\tau_{5D_{5/2}})+A_2\exp(-t/\tau_{6P_{3/2}})+B,
\end{equation}
where $\tau_{5D_{5/2}}$ and $\tau_{6P_{3/2}}$ are the lifetimes of
the $5D_{5/2}$ and $6P_{3/2}$ states, respectively and B is a
possible background.

If we leave both lifetimes as free parameters, the L-M algorithm
does not give reliable results (it tends to equate both
lifetimes). Therefore, we fix the value of $\tau_{6P_{3/2}}$ in
the fitting process to solve this problem, and the value we use is
the weighted average $\tau_{6P_{3/2}}$=112.8$\pm1.7$~ns from
previous experimental work listed in Ref.~\cite{marek80}. Our
fitting process gives an averaged reduced $\chi^2$ of 1.09 for all
the data sets we have taken. The lower plot of
Fig.~\ref{fig:counts5D25} shows the normalized fit residuals by
comparing the raw data and fit result in the upper plot.

\subsection{Statistical error}
We have taken 10 sets of data with the same experimental
conditions, with each individual data set fitting error,
$\sigma_i$, of around 1\%. Following the same procedure in the
Sec.~\ref{sec:5d23},  we get a weighted mean lifetime of 238.6~ns,
error of the mean as 0.8~ns, and the scaling factor $S=1.2$. We
choose the scaled error of the mean as the statistical error,
which is 1.0~ns.

\subsection{Systematic effects}
We have studied all systematic effects listed in the measurement
of the $5D_{3/2}$ state, which shows similar results. We focus on
the following new situations in this measurement:

\emph{Quantum beats.} It is possible to coherently excite the
atoms to the Zeeman sublevels of the $5D_{5/2}$ state, but to
observe the quantum beats signal for this cascade decay, we also
need to prepare the coherence on the $6P_{3/2}$ state, for
example, by means of selecting polarization on the fluorescence
\cite{aspect84}. In our experiment, we do not attempt to observe
the coherence in the intermediate level, and do not expect
observable quantum beats effects. We check the Fourier transform
of the fitting residuals and do not find peaks. We also check the
magnetic field effects on the lifetime measurement and do not see
any systematic effects. We set the error as 0.44\%, which is
defined in same way of that in the systematic study in
Sec.~\ref{sec:5d23}.

\emph{Bayesian error.} In the fitting process, we do not treat
$\tau_{6P_{3/2}}$ and $\tau_{5D_{5/2}}$ as independent parameters
because we fix the $\tau_{6P_{3/2}}$ by its experimental value.
Hence there is correlation between the values of $\tau_{6P_{3/2}}$
and $\tau_{5D_{5/2}}$ in the fitting process. We express the
probability of finding $\tau_{5D_{5/2}}$ equal to $\tau_1'$ as
\begin{equation}
P(\tau_1')={\int}P(\tau_1'|\tau_2')P(\tau_2')d\tau_2',
\label{eq:baye}
\end{equation}
where $\tau_2'$ denotes the possible value of $\tau_{6P_{3/2}}$
and $P(\tau_1'|\tau_2')$ is the conditional probability of getting
$\tau_{5D_{5/2}}=\tau_1'$ when $\tau_{6P_{3/2}}=\tau_2'$.
Eq.~(\ref{eq:baye}) shows that the error of $\tau_{6P_{3/2}}$
value propagates to that of $\tau_{5D_{5/2}}$ value through the
correlation, and this is the Bayesian error.

Following the treatment of Ref.~\cite{aubin04}, we change the
value of $\tau_{6P_{3/2}}$ from 108~ns to 118~ns with an increase
of 0.5~ns each step, and do the fitting process to extract a new
$\tau_{5D_{5/2}}$ for each new value of $\tau_{6P_{3/2}}$. This
scanning range has covered 3$\sigma$ range of the experimental
result of $\tau_{6P_{3/2}}$. We find that a linear function
describes well the dependance of the fitting result of
$\tau_{5D_{5/2}}$ on the value of $\tau_{6P_{3/2}}$
\begin{equation}
\tau_1(\tau_2)=\tau_1+\alpha(\tau_2'-\tau_2),
\end{equation}
where the notation is the same as in Eq.~(\ref{eq:baye}). Hence we
get the Bayesian error as $|\alpha|\sigma_{\tau_{6P_{3/2}}}$. For
our data sets, the analysis gives $\alpha$ as -0.3, and the
Bayesian error as 0.5~ns.

\emph{Other systematics.} We study the pulse length and power
effects, and do not find any systematic changes in the lifetime
values. The limit set on this error due to these effects is
0.73\%.

\subsection{Summary and Comparison}

We list the error budget of the $5D_{5/2}$ state lifetime
measurement in the table \ref{tab:err5d25} and compare our result
with the previous experimental and theoretical work in table
\ref{tab:ret5d25}. The result of Ref. \cite{safronova04} sits
within the $2\sigma$ range of our data, and that of Ref.
\cite{theodosiou84} sits close to the edge of the $3\sigma$ range.
Since the error of the all-order method estimation on this level
is 5\% (see the Sec.~\ref{subsec:redsum}), the result of
Ref.~\cite{safronova04} is consistent with our result. On the
experimental part, as far as we know, there has been no explicit
experiment data on the $5D_{5/2}$ state lifetime.

The measurement of Ref. \cite{marek80} gave a lifetime of the
combination of $5D_{3/2}$ and $5D_{5/2}$ states. According to our
results, the lifetime of the whole $5d$ manifold should sit
between 238.5~ns and 246.3~ns, which is consistent with the result
of Ref. \cite{marek80}.

\begin{table}
\caption{\label{tab:err5d25} Error budget of the $5D_{5/2}$ state
lifetime measurement.}
\begin{ruledtabular}
\begin{tabular}{lcc}
Source&Correction (\%)&Error(\%)\\
\hline
Statistical&&$\pm$0.40\\
Time calibration&&$<\pm$0.01\\
TDC nonuniformity&&$\pm$0.01\\
Pulse pileup&-0.05&\\
Magnetic filed&&$\pm$0.44\\
Radiation trapping&&$<\pm$0.1\\
Bayesian&&$\pm$0.21\\
Other Systematics&&$\pm0.73$\\
\\
\textbf{Total}&&$\pm$\textbf{0.96}
\end{tabular}
\end{ruledtabular}
\end{table}

\begin{table}
\caption{\label{tab:ret5d25} Comparison of the measured lifetime
of the $5D_{5/2}$ state with the previous work.}
\begin{ruledtabular}
\begin{tabular}{llc}
&& $\tau_{5D_{3/2}}$ (ns)\\
\hline
Experiment&\textbf{This work}&\textbf{238.5$\pm$2.3}\\
&Marek and Munster \footnote{This measurement gave a value for the
whole 5d manifold.} \cite{marek80}&230$\pm$23\\
\\
Theory&Theodosiou \cite{theodosiou84}&231\\
&Safronova et al. \cite{safronova04}&235
\end{tabular}
\end{ruledtabular}
\end{table}

\section{Conclusion\label{sec:end}}
In this work, we have measured the lifetime of the Rb $5D_{5/2}$
state obtaining $\tau_{5D_{3/2}}=246.3\pm1.6$~ns, and we have
improved the precision of the $5D_{3/2}$ lifetime result,
$\tau_{5D_{5/2}}=238.5\pm2.3$~ns, by a factor of 25. Our results
have enough precision to confirm the improvement of the scaled
all-order method \cite{safronova04}.

Our experimental condition present several advantages over
previous experimental efforts: the new laser technique which
allows us to optically excite the atoms to the desired states
directly, the lock technique of both lasers in the two-step
transition which makes the transition much more efficient and the
use of a MOT which simplifies the systematic studies.

Our future work, the search of anapole moment in alkali atoms
\cite{gomez07}, relies on high precision atomic structure
calculations to extract the fundamental information from the
measurement. We hope the advance on the experimental value of the
lifetime will contribute to new improvements on the precision of
theoretical calculations.

\begin{acknowledgements}
This work is supported by the NSF. We thank C. Monroe and S.
Rolston for support with equipment, and M. S. Safronova for
helpful discussions.
\end{acknowledgements}


\end{document}